\documentclass[12pt]{article}

\usepackage{sbc-template}
\usepackage{graphicx,url}
\usepackage{hyperref}

\usepackage[latin1]{inputenc}  
\usepackage{multirow}  
\usepackage[usenames,dvipsnames]{color}
\usepackage[usenames,dvipsnames]{xcolor}
\usepackage{float}
\usepackage{listings}  
\usepackage{color}
\usepackage{amsfonts}
\usepackage{subfigure}
\usepackage{colortbl}
\usepackage{booktabs}
\usepackage{amsmath}
\usepackage{algorithmicx}
\usepackage{algorithm}
\usepackage{algpseudocode}

\mathchardef\mhyphen="2D
     
\newcommand*{\var}[1]{%
$\mathtt{#1}$%
}

\newcommand*{\vart}[1]{%
\texttt{\textbf{#1}}%
}

\definecolor{javagreen}{rgb}{0.25,0.5,0.35}
\definecolor{javapurple}{rgb}{0.5,0,0.35}
\definecolor{javadocblue}{rgb}{0.25,0.35,0.75}
\title{JExtract: An Eclipse Plug-in for Recommending Automated Extract Method Refactorings}

\author{Danilo Silva\inst{1}, Ricardo Terra\inst{2}, Marco Túlio Valente\inst{1}}

\address{Federal University of Minas Gerais, Brazil
\nextinstitute
  Federal University of Lavras, Brazil
  \email{\footnotesize \{danilofs,mtov\}@dcc.ufmg.br, terra@dcc.ufla.br}
}

\begin{document} 

\maketitle

\begin{abstract}
Although Extract Method is a key refactoring for improving program comprehension,
refactoring tools for such purpose are often underused.
To address this shortcoming, we present JExtract, a recommendation system based on structural similarity 
that identifies Extract Method refactoring opportunities 
that are directly automated by IDE-based refactoring tools.
Our evaluation
suggests that JExtract 
is more effective (w.r.t.~recall and precision) to identify contiguous misplaced code in methods than JDeodorant, 
a state-of-the-art tool.

\vspace{10pt}
\noindent {\textbf{Tool demonstration video.} \href{http://youtu.be/6htJOzXwRNA}{http://youtu.be/6htJOzXwRNA}}

\end{abstract}

\section{{Introduction}}

Refactoring has increased in importance as a technique for improving the design of \mbox{existing}
code~\cite{fowler99}, e.g., to increase cohesion, decrease coupling, foster maintainability, etc. 
Particularly, Extract Method is a key refactoring for improving program comprehension. 
Besides promoting reuse and reducing code duplication, it contributes to readability and comprehensibility, by encouraging the extraction of self-documenting methods~\cite{fowler99}. 

Nevertheless, recent empirical research indicate that, while Extract Method is one of the most common refactorings, automated tools supporting this refactoring are most of the times underused~\cite{negara2013,MurphyHill2012}.
For example, Negara et al.~found that Extract Method is the third most frequent refactoring, but the number of developers who apply the refactoring manually is higher than the number of those who do it automatically~\cite{negara2013}.
Moreover, current tools focus only on automating refactoring application, but developers expend considerable effort on the manual identification of refactoring opportunities.

%

To address this shortcoming, this paper presents \var{JExtract},
a tool that implements a novel approach for recommending automated Extract Method refactorings.
The tool was designed as a plug-in for the Eclipse IDE that automatically identifies, ranks, and applies the refactoring when requested. Thereupon, \var{JExtract} may aid developers to find refactoring opportunities and contribute to a widespread adoption of refactoring practices.
The underlying technique
is inspired by the {\em separation of concerns} design guideline. More specifically, we assume that {\em the structural dependencies established by Extract Method candidates should be very different from the ones established by the remaining statements in the original method}. 

    The remainder of this paper is structured as follows. 
    Section~\ref{sec:tool} describes the $\tt JExtract$ tool, including its design and implementation.
    Section~\ref{sec:relatedtools} discusses related tools and 
    Section~\ref{sec:finalremarks} presents final remarks.
    

\section{{The \vart{JExtract} tool}}
\label{sec:tool}



\begin{figure}[htpb]
\centering
\includegraphics[width=1.0\textwidth]{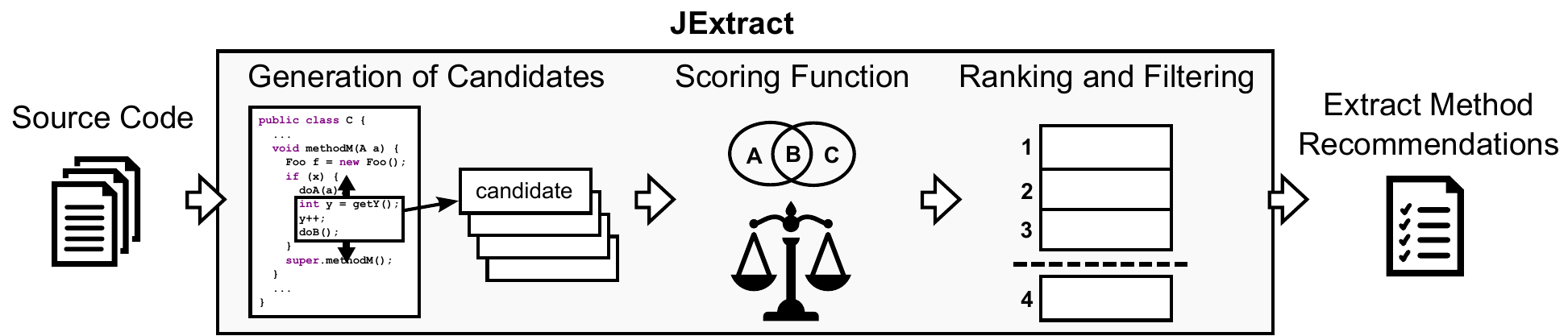}
\vspace{-19pt}
\caption{The \vart{JExtract} tool}
\label{image_approachoverview}
\end{figure}

\var{JExtract} is a tool that analyzes the source code of methods and recommends Extract Method refactoring opportunities, as illustrated in Figure~\ref{image_approachoverview}. First, the tool generates all Extract Method possibilities for each method. Second, these possibilities are ranked according to a scoring function based on the similarity between sets of dependencies established in the code.

This main section of the paper is organized as follows.
Subsection~\ref{sec:approach} provides an overview of our approach for identifying
Extract Method refactoring opportunities. Subsection~\ref{sec:archinterface} describes the design and implementation of the tool. Finally,
Subsection~\ref{sec:examples} presents the results of our evaluation in
open-source systems. A detailed description of the recommendation technique behind JExtract is present in a recent full technical paper~\cite{2014_icpc}.

\subsection{{Proposed Approach}} 
\label{sec:approach}

The approach is divided in three phases: {\em Generation of Candidates}, {\em Scoring}, and {\em Ranking}. 

\subsubsection{{Generation of candidates}} 
\label{sec:candidate_generation}

\begin{figure}[H]
\centering
\includegraphics[width=0.80\textwidth]{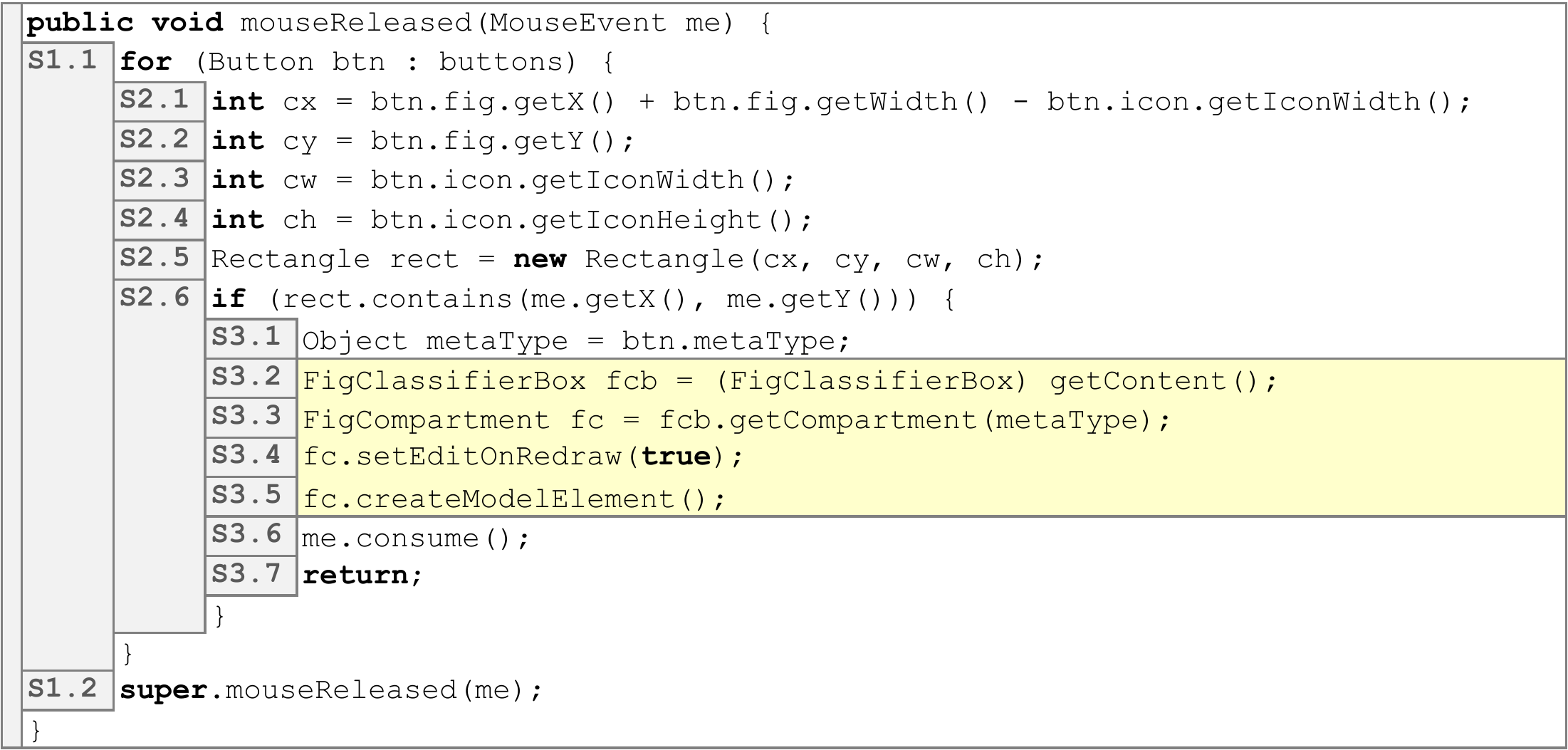}
\vspace{-6pt}
\caption{An Extract Method candidate in a method of \mbox{ArgoUML} (\vart{S3.2} to \vart{S3.5})}
\label{image_example1latex}
\vspace{-6pt}
\end{figure}

This phase is responsible for identifying all possible Extract Method refactoring opportunities. First, we split the methods into blocks, which consist of sequential statements that follow a linear control flow. As an example, Figure~\ref{image_example1latex} presents method \var{mouseRelease} of class \var{SelectionClassifierBox}, extracted from ArgoUML. We can notice that each statement is labeled using the \var{SX.Y} pattern, where $\tt{X}$ and $\tt{Y}$ denote the block and the statement, respectively. For example, $\tt S2.3$ is the third statement of the second block, which declares a variable $\tt{cw}$.

%


Second, we generate all Extract Method candidates based on Algorithm~\ref{algo:generation} (extracted from~\cite{2014_icpc}).

\vspace{-6pt}
\begin{algorithm}[!ht]
{\scriptsize
\caption{Candidates generation algorithm~\cite{2014_icpc}}
{\textbf{Input:} A method $M$\\}
{\textbf{Output:} List with Extract Method candidates}
\begin{algorithmic}[1]
\State $Candidates \gets \emptyset$
\ForAll{$block\ B \in M$}
  \State $n \gets statements(B)$
  \For{$i \gets 1, n$}
    \For{$j \gets i, n$}
      \State $C \gets subset(B, i, j)$
      \If{$isValid(C)$}
        \State $Candidates \gets Candidates + C$
      \EndIf
    \EndFor
  \EndFor
\EndFor
\end{algorithmic}
\label{algo:generation}
}
\end{algorithm}
\vspace{-20pt}

\vspace*{0.4cm}
\mbox{Fundamentally}, the three nested loops in Algorithm~\ref{algo:generation} (lines~2, 4, and 5) enforce that the list of selected statements attend the following preconditions:
%

\begin{itemize}
\item Only continuous statements inside a block are selected. In Figure~\ref{image_example1latex}, for example, it is not possible to select a candidate with $\tt S3.2$ and $\tt S3.4$ without including $\tt S3.3$.\\[-0.3cm]

\item The selected statements are part of a single block of statements. In Figure~\ref{image_example1latex}, for example, it is not possible to generate a candidate with both $\tt S2.6$ and $\tt S3.1$ since they belong to distinct blocks.\\[-0.3cm]

\item When a statement is selected, the respective children statements are also included. In Figure~\ref{image_example1latex}, for example, when statement $\tt S2.6$ is selected, its 
children statements $\tt S3.1$ to $\tt S3.7$ are also included.
\end{itemize}

Last but not least, we do not ensure that every iteration of the loop yields an Extract Method candidate because: (i)~a candidate recommendation must respect a size threshold defined by parameter \textit{Minimum Extracted Statements}. The value is preset to $3$ {\small (changeable)},
which means that an Extract Method candidate has to have at least three statements; and (ii)~a candidate recommendation must respect the preconditions defined by the Extract Method refactoring engine.

\subsubsection{{Scoring}}
\label{sec:scoring_function}

%
%
This phase is responsible for scoring the possible Extract Method refactoring opportunities generated in the previous phase, using a technique inspired by a Move Method recommendation heuristic~\cite{jmove}.
Assume $m'$ as the selection of statements of an Extract Method candidate and $m''$ the remaining statements in the original method~$m$.
The proposed heuristic aims to minimize the structural similarity between $m'$ and~$m''$.
\vspace{6pt}

\noindent{\bf Structural Dependencies:} The set of dependencies established by a selection of statements $S$ with variables, types, and packages is denoted by $\mathit{Dep}_{\mathit{var}}(S)$, $\mathit{Dep}_{\mathit{type}}(S)$, and $\mathit{Dep}_{\mathit{pack}}(S)$, respectively. These sets are constructed as described next.

\vspace{3pt}

\begin{itemize}
\item{\em Variables:} If a statement $s$ from a selection of statements $S$ declares, assigns, or reads a variable $v$, then $v$ is added to $\mathit{Dep}_{\mathit{var}}(S)$. In a similar way, reads from and writes to formal parameters and fields are considered.\\[-0.2cm]

\item{\em Types:} If a statement $s$ from a selection of statements $S$ uses a type (class or interface) $T$, then $T$ is added to $\mathit{Dep}_{\mathit{type}}(S)$.\\[-0.2cm] 

\item{\em Packages:} For each type $T$ included in $\mathit{Dep}_{\mathit{type}}(S)$, as described in the previous item, the package where $T$ is implemented and all its parent packages are also included in $\mathit{Dep}_{\mathit{pack}}(S)$.

\end{itemize}
\vspace{3pt}

For instance, assume $m'$ as the highlighted code in Figure~\ref{image_example1latex} (i.e., an Extract Method candidate) and $m''$ the remaining statements in the original method~\var{mouseReleased}.
On one hand,
$\mathit{Dep}_{\mathit{var}}(m') = \left\{ {\tt{metaType}, \tt{fc}, \tt{fcb}}\right\}$.
On the other hand, the set $\mathit{Dep}_{\mathit{var}}(m'') = \left\{ {\tt{metaType}, \tt{btn}, \tt{cy}, \tt{cx}, \tt{cw}, \tt{ch}, \tt{buttons}, \tt{me}, \tt{rect}}\right\}$.
In this case, the intersection between these two sets contains only ${\tt{metaType}}$. Moreover, the computation of ${\tt fc}$ and ${\tt fcb}$ is isolated from the remaining code. Therefore, one can claim that $m'$ is cohesive and decoupled from $m''$, i.e.,~a good separation of concerns is achieved. \\



\noindent{\bf Scoring Function}: To compute the dissimilarity between $m'$ and $m''$, we rely on the distance between the dependency sets $Dep'$ and $Dep''$ using the Kulczynski similarity coefficient~\cite{terra-seke-13,jmove}:
\[
dist(Dep',Dep'') = 1 - \frac{1}{2}\Big[\frac{a}{(a+b)}+\frac{a}{(a+c)}\Big]
\]
\noindent where  ${a}= |Dep'\, \bigcap\, Dep''|$,  ${b}= |Dep' \setminus Dep''|$, and \mbox{${c}= | Dep'' \setminus Dep'|$}.\\

Thus, let $m'$ be the selection of statements of an Extract Method candidate for method $m$. Let also $m''$ be the remaining statements in $m$. The score of $m'$ is defined as:\\

$
\begin{array}{ll}
\mathit{score}(m')  = & 1/3 \times \mathit{dist}(\mathit{Dep}_{\mathit{var}}(m'),\mathit{Dep}_{\mathit{var}}(m''))\,\,\, + \\[.1cm]
                     & 1/3 \times \mathit{dist}(\mathit{Dep}_{\mathit{type}}(m'),\mathit{Dep}_{\mathit{type}}(m''))\,\,\, + \\[.1cm]
                     & 1/3 \times \mathit{dist}(\mathit{Dep}_{\mathit{pack}}(m'),\mathit{Dep}_{\mathit{pack}}(m'')) \\[.1cm]
\end{array}
$\\

The scoring function is centered on the observation that 
a good Extract Method candidate should encapsulate the use of variables, types, and packages. In other words, we should maximize the distance between the dependency sets $Dep'$ and $Dep''$.

\subsubsection{{Ranking}}
This phase is responsible for ranking and filtering the Extract Method candidates
based on the score computed in the previous phase.
Basically, we sort the candidates and filter them according to the following parameters: (i)~{\em Maximum Recommendations per Method}. The value is preset to 3 {\small (changeable)}, which means that the tool triggers up to three recommendations for each method; and (ii)~{\em Minimum Score Value}, which has to be configured when the user desires
to setup a minimum dissimilarity threshold. 

\subsection{{Internal Architecture and Interface}}
\label{sec:archinterface}

We implemented \var{JExtract} as a plug-in on top of the Eclipse platform. 
Therefore, we rely mainly on native Eclipse APIs, such as Java Development Tools (JDT)
and Language Toolkit (LTK). 
The current \var{JExtract} implementation follows an architecture with five main modules:\\[-0.1cm]



\begin{enumerate}

\item {\em Code Analyzer}: This module provides the following services to other
modules: (a)~it builds the structure of block and statements {\small (refer to Subsection~\ref{sec:candidate_generation})}; (b)~it extracts the structural dependencies
{\small (refer to Subsection~\ref{sec:scoring_function})}; and (c)~it checks if an Extract
Method candidate satisfies the underlying Eclipse Extract Method refactoring preconditions.
In fact, this module contains most communication between \var{JExtract}
and Eclipse APIs (e.g., \var{org.eclipse.jdt.core} and \var{org.eclipse.ltk.core.refactoring}). \\[-0.1cm]


\item {\em Candidate Generator}: This module generates all Extract Method candidates based on Algorithm~\ref{algo:generation} and hence depends on service (a) of module {\em Code Analyzer}.\\[-0.1cm]

\item {\em Scorer}: This module calculates the dissimilarity of the Extract Method candidates generated by module {\em Candidate Generator} {\small (refer to Subsection~\ref{sec:scoring_function})} and hence
depends on service (b) of module {\em Code Analyzer}.\\[-0.1cm]

\item {\em Ranker}:
This module ranks and filters the Extract Method candidates generated by  
module {\em Candidate Generator} and scored by module {\em Scorer}. It depends
on service~(c) of module {\em Code Analyzer} to filter candidates not satisfying preconditions.
\\[-0.1cm]

\item {\em UI}: This module consists of the front-end of the tool, which relies on the Eclipse UI API (\var{org.eclipse.ui}) to implement two menu extensions, six actions, and one main view.
Moreover, it depends on module UI from LTK (\var{org.eclipse.ltk.ui.refactoring}) to
delegate the refactoring appliance to the underlying Eclipse Extract Method refactoring tool.\\[-0.1cm]

%
%
%

\end{enumerate}

Such architecture permits the extension of our tool. For example, the {\em Scorer} module may be replaced by one that employs a new heuristic based on semantic and structural information. As another example, the {\em Candidate Generator} module may be extended to support the identification of non-contiguous code fragments.

\begin{figure}[htpb]
\vspace{-6pt}
\centering
\includegraphics[width=0.8\textwidth]{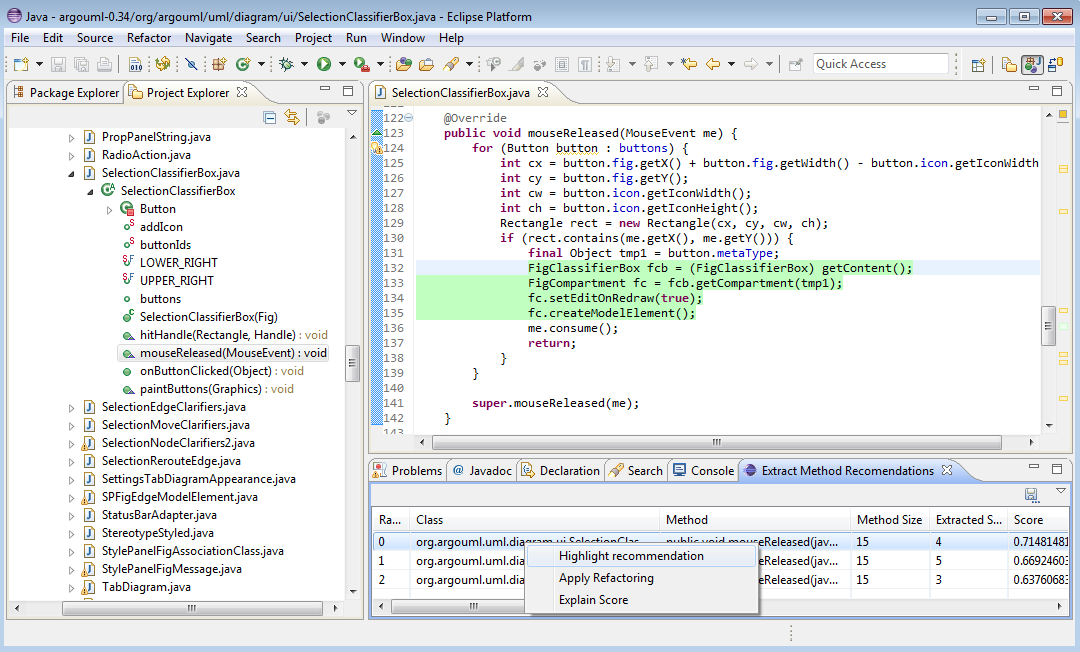}
\caption{JExtract UI}
\label{img_uilatex}
\vspace{-6pt}
\end{figure}

Figure~\ref{img_uilatex} presents \var{JExtract}'s UI, displaying method \var{mouseReleased} previously presented in Figure~\ref{image_example1latex}. When a developer triggers \var{JExtract} to identify Extract Method refactoring opportunities for this method, it opens the {\em Extract Method Recommendations} view to report the potential recommendations. In this case, the best candidate consists of the extraction of statements $\tt S3.2$ to $\tt S3.5$ whose dissimilarity score is 0.7148.



\subsection{{Evaluation}}
\label{sec:examples}

We conducted two different but complementary empirical studies.

\vspace{5pt}
\noindent{\bf Study \#1}:
In our previous paper~\cite{2014_icpc}, we evaluated the recommendations provided by our tool on three systems to assess precision and recall. 
We extended this study to consider minor modifications to the ranking method and to compare the results with \var{JDeodorant}, a state-of-the-art tool that identifies Extract Method opportunities~\cite{tsantalis11}.
For each system~$S$, we apply random Inline Method refactoring operations to obtain a modified version~$S'$. 
%
%
We assume that good Extract Method opportunities are the ones that revert the modifications (i.e., restoring $S$ from $S'$).




\begin{table}[H]
\vspace{-6pt}
\caption{Study \#1 -- Recall and precision results} 
\begin{center}
{\footnotesize
\begin{tabular}{lr|rr|rr|rr|rr}
\hline
& & \multicolumn{6}{c|}{\vart{JExtract}} & \multicolumn{2}{c}{\vart{JDeodorant}} \\
 & & \multicolumn{2}{c}{\bf Top-1} & \multicolumn{2}{c}{\bf Top-2} & \multicolumn{2}{c|}{\bf Top-3} &  \\
{\bf System} & \multicolumn{1}{c|}{\bf\#} & \multicolumn{1}{c}{\bf Recall} & \multicolumn{1}{c}{\bf Prec.}  & \multicolumn{1}{c}{\bf Recall} & \multicolumn{1}{c}{\bf Prec.}  & \multicolumn{1}{c}{\bf Recall} & \multicolumn{1}{c|}{\bf Prec.} & \multicolumn{1}{c}{\bf Recall} & \multicolumn{1}{c}{\bf Prec.} \\
\hline
JHotDraw~5.2 & 56 & 19 {\scriptsize(34\%)} & 34\% & 26 {\scriptsize(46\%)} & 24\% & 32 {\scriptsize(57\%)} & 20\% & 2 {\scriptsize(4\%)} & 5\% \\
JUnit~3.8 & 25 & 13 {\scriptsize(52\%)} & 52\% & 16 {\scriptsize(64\%)} & 33\% & 18 {\scriptsize(72\%)} & 25\% & 0 {\scriptsize(0\%)} & 0\% \\
MyWebMarket & 14 & 12 {\scriptsize(86\%)} & 86\% & 14 {\scriptsize(100\%)} & 50\% & 14 {\scriptsize(100\%)} & 33\% & 2 {\scriptsize(14\%)} & 33\% \\
\hline
{\bf\em Total} & {\bf\em 95} & {\bf\em 44 {\scriptsize(46\%)}} & 46\% & {\bf\em 56 {\scriptsize(59\%)}} & {\bf\em 30\%} & {\bf\em 64 {\scriptsize(67\%)}} & {\bf\em 23\%} & {\bf\em 4 {\scriptsize(4\%)}} & {\bf\em 6\%} \\
\hline
\end{tabular}
\label{table:evaluation1}
}
\end{center}
\vspace{-6pt}
\end{table}

Table~\ref{table:evaluation1} reports recall and precision values achieved using \var{JExtract} with three different configurations {\small ({\em Top-k Recommendations per Method})}. 
While a high parameter value favors recall {\small (e.g., Top-3)}, a low one favors precision {\small (e.g., Top-1)}. Table~\ref{table:evaluation1} also presents results achieved using \var{JDeodorant} with its default settings.
As the main finding, \var{JExtract} outperforms \var{JDeodorant} regardless of the configuration used.


\vspace{5pt}
\noindent{\bf Study \#2}:
%
%
We replicate the previous study in other ten popular open-source Java systems
to assess how the precision and recall rates would vary. Nevertheless, we do not compare our results with \var{JDeodorant} since we were not able to reliably provide the source code of all required libraries, as demanded by JDeodorant.



\begin{table}[htbp]
\vspace{-6pt}
\caption{Study \#2 -- Recall and precision results}
\begin{center}
{\footnotesize
\begin{tabular}{l@{\hspace{3pt}}r|rr|rr|rr}
\hline
& & \multicolumn{6}{c}{\vart{JExtract}} \\
 & & \multicolumn{2}{c}{\bf Top-1} & \multicolumn{2}{c}{\bf Top-2} & \multicolumn{2}{c}{\bf Top-3}  \\
{\bf System} & \multicolumn{1}{c|}{\bf \#} & \multicolumn{1}{c}{\bf Recall} & \multicolumn{1}{c}{\bf Prec.}  & \multicolumn{1}{c}{\bf Recall} & \multicolumn{1}{c}{\bf Prec.}  & \multicolumn{1}{c}{\bf Recall} & \multicolumn{1}{c}{\bf Prec.} \\
\hline
Ant~1.8.2 & 964 & 235 {\scriptsize (24.4\%)} & 24.4\%  & 363 {\scriptsize (37.7\%)} & 19.1\% & 460 {\scriptsize (47.7\%)} & 16.3\% \\
ArgoUML~0.34 & 439 & 98 {\scriptsize (22.3\%)} & 22.3\%  & 160 {\scriptsize (36.4\%)} & 18.3\% & 186 {\scriptsize (42.4\%)} & 14.4\% \\
Checkstyle~5.6 & 533 & 227 {\scriptsize (42.6\%)} & 42.6\%  & 338 {\scriptsize (63.4\%)} & 31.9\% & 389 {\scriptsize (73.0\%)} & 24.7\% \\
FindBugs~1.3.9 & 714 & 179 {\scriptsize (25.1\%)} & 25.1\%  & 278 {\scriptsize (38.9\%)} & 19.7\% & 350 {\scriptsize (49.0\%)} & 16.7\% \\
FreeMind~0.9.0 & 348 & 85 {\scriptsize (24.4\%)} & 24.4\%  & 134 {\scriptsize (38.5\%)} & 19.4\% & 181 {\scriptsize (52.0\%)} & 17.8\% \\
JFreeChart~1.0.13 & 1,090 & 204 {\scriptsize (18.7\%)} & 18.7\%  & 396 {\scriptsize (36.3\%)} & 18.2\% & 536 {\scriptsize (49.2\%)} & 16.5\% \\
JUnit~4.10 & 35 & 11 {\scriptsize (31.4\%)} & 32.4\%  &  17 {\scriptsize (48.6\%)} & 26.6\% &  22 {\scriptsize (62.9\%)} & 23.7\% \\
Quartz~1.8.3 & 239 & 99 {\scriptsize (41.4\%)} & 41.4\%  & 125 {\scriptsize (52.3\%)} & 26.5\% & 142 {\scriptsize (59.4\%)} & 20.4\% \\
SQuirreL~SQL~3.1.2 & 39 & 15 {\scriptsize (38.5\%)} & 38.5\%  &  18 {\scriptsize (46.2\%)} & 23.7\% &  20 {\scriptsize (51.3\%)} & 18.2\% \\
Tomcat~7.0.2 & 1,076 & 214 {\scriptsize (19.9\%)} & 19.9\%  & 325 {\scriptsize (30.2\%)} & 15.2\% & 409 {\scriptsize (38.0\%)} & 12.8\% \\
\hline
{\bf\em Total} & {\bf\em 5,477} & {\bf\em 1,367 {\scriptsize (25.0\%)}} & {\bf\em 25.0\%} & {\bf\em 2,154 {\scriptsize (39.3\%)}} & {\bf\em 19.8\%} & {\bf\em 2,695 {\scriptsize (49.2\%)}} & {\bf\em 16.7\%} \\
\hline
\end{tabular}
\label{table:evaluation2}
}
\end{center}
\vspace{-6pt}
\end{table}

Table~\ref{table:evaluation2} reports the 
recall and precision values achieved using the same settings from the previous study. 
On one hand, the overall recall value ranges from 25\% to 49.2\%. On the other hand, the overall precision value ranges from 25\% to 16.7\%. 
We argue these values are acceptable for two reasons: (i)~we only consider as correct a recommendation that matches exactly the one at the oracle; thus, a slight difference of including (or excluding) a statement is enough to be considered a miss; and (ii) the modified methods may have preexisting Extract Method opportunities, besides the ones we introduced, that will be considered wrong by our oracle.%



\section{{Related Tools}}
\label{sec:relatedtools}
Recent empirical research shows that automated refactoring tools, especially those supporting Extract Method refactorings, are most of the times underused~\cite{negara2013,MurphyHill2012}. 
In view of such circumstances, recent studies on identification of refactoring opportunities
are seeking to address this shortcoming. In this paper,  
%
%
we implemented our approach  
in a way that it can be straightforwardly
incorporated to the current development process through a tool 
that identifies, ranks, and automate Extract Method refactoring opportunities~\cite{2014_icpc}.

 \var{JMove} is the refactoring recommendation system our approach is inspired by~\cite{jmove,2013_cbsoft_tool}. The tool identifies Move Method refactoring opportunities based on the similarity between dependency sets~\cite{jmove}. More specifically,
 it computes the similarity of the set of dependencies 
 established by a given method~$m$ with (i)~the methods of its own class~$C_1$
 and (ii) the methods in other classes of the system {\small ($C_2,C_3,...,C_{n}$)}. 
  %
%
%
Whereas JMove recommends moving a method~$m$ to a more similar class~$C_i$, our current approach recommends extracting a fragment from a given method~$m$ into a new method~$m'$ when there is a high dissimilarity between $m'$ and the remainder statements in~$m$.



\var{JDeodorant} is the state-of-the-art system 
to identify and apply common refactoring operations in Java systems, including Extract Method~\cite{tsantalis11}.
In contrast to our approach that relies on the similarity between dependency sets, 
\var{JDeodorant} relies on the concept of program slicing to select related statements that can be extracted into a new method. 
Our approach, on the other hand, is not based on specific code patterns (such as a computation slice). It is also more conservative to preserve program behavior (although it is currently restricted to non-contiguous fragments of code), and it relies on a scoring function to rank and filter recommendations.

There are other techniques to identify refactoring opportunities based, for example, on search-based algorithms~\cite{Seng2006}, Relational Topic Model (RTM)~\cite{methodbook_tse},  metrics-based rules~\cite{Marinescu:2004:DSM:1018431.1021443}, etc., that can be adapted to identify Extract Method refactoring opportunities.
%

\section{{Final Remarks}}
\label{sec:finalremarks}

\var{JExtract} implements a novel approach for recommending automated Extract Method refactorings.  
The tool 
was designed as a plug-in for the Eclipse IDE that automatically identifies, ranks, and applies the refactoring.
Thereupon, the tool may contribute to increase the popularity of IDE-based 
refactoring tools, which are normally considered underused by most recent empirical studies on refactoring. 
Moreover, our evaluation indicates that \var{JExtract} is more effective (w.r.t.~recall and precision) to identify contiguous misplaced code in methods than \var{JDeodorant}, a state-of-the-art tool.



As ongoing work, we are extending \var{JExtract} 
to be able to do statement reordering to uncover better Extract Method opportunities, as long as the modification preserves the behavior of the original code. 
Moreover, we intend to evaluate our tool with human experts to mitigate the threat that the synthesized datasets did not capture the full spectrum of Extract Method instances faced by developers.
Last, we also intend to support other kinds of refactoring (e.g., Move Method).


The \var{JExtract} tool---including its source code---is publicly
available at \href{http://aserg.labsoft.dcc.ufmg.br/jextract}{$\tt http{:}//aserg.labsoft.dcc.ufmg.br/jextract$}.%

\noindent{\bf Acknowledgments}:~Our research is supported by CAPES, FA\-PE\-MIG, and CNPq. 

\linespread{0.78}
\footnotesize
\bibliographystyle{abbrv}
\bibliography{terracopy}

\end{document}